\documentclass{elsart}
\usepackage{psfig}

\begin{document}

\begin{frontmatter}

\title{Possible Discrimination between Gamma Rays and Hadrons 
using \v Cerenkov Photon Timing Measurements}

\author{V. R. Chitnis and P. N. Bhat}
\address{Tata Institute of Fundamental Research,
 Homi Bhabha Road,
 Mumbai 400 005, India.}

\begin{abstract}
Atmospheric \v Cerenkov Technique is an established methodology to study
$TeV$ energy gamma rays. However the challenging problem has always been the
poor signal to noise ratio due to the presence of abundant cosmic rays. Several
ingenious techniques have been
employed to alleviate this problem, most of which are centred around the \v
Cerenkov image characteristics. However there are not many techniques
available for improving the signal to noise ratio of the data from 
wavefront sampling observations.  One such possible technique is to use
the \v Cerenkov photon arrival times and identify the species dependent
characteristics in them.  Here we carry out systematic monte carlo
simulation studies of the timing information of \v Cerenkov photons at the
observation level. We have parameterized the shape of the \v Cerenkov
shower front as well as the pulse shapes in terms of experimentally
measurable quantities. 
We demonstrate the sensitivity of the curvature of the shower front, 
pulse shape parameters as well as the photon arrival time jitter to
primary species and show their efficiency in improving the signal to noise
ratio. The effect of limiting the \v Cerenkov telescope opening angle by 
using a circular focal point mask, on the efficacy of the parameters has also been 
studied for each of the parameters. Radius of the shower front, pulse decay 
time and photon arrival time jitter have been found to be the most
promising parameters which could be used to discriminate $\gamma -$ray
events from the background. We also find that the efficiency of the first two
parameters increases with zenith angle and efficiency of pulse decay time 
decreases with increasing altitude of observation. 

\end{abstract}
\end{frontmatter}

\section{Introduction}

Atmospheric \v Cerenkov Technique (ACT), is a well established and
a unique method for the astronomical investigation of Very High Energy
(VHE, also referred to as $TeV$) $\gamma -$ rays. It is based on the
effective detection and study of the \v Cerenkov light emitted by the
secondary particles produced in the extensive air showers initiated by the
primary $\gamma -$ ray \cite{cr93,fe97,sc96,on98}.

As is well known, the abundant charged cosmic ray particles generate
\v Cerenkov light akin to that produced by the $\gamma -$ rays as a result
of which the $\gamma -$ ray signal is buried in a vast sea of cosmic ray
background. This has been a major difficulty in applying the atmospheric
\v Cerenkov technique successfully for $\gamma -$ ray astronomy.

The first generation \v Cerenkov telescopes addressed the problem of cosmic
ray background by matching the telescope aperture to the angular size of
the \v Cerenkov flash \cite{br89}. In addition, many of the potential astrophysical
sources of VHE $\gamma -$ rays are expected to produce modulated signals
({\it e.g.} $\gamma -$ ray pulsars, X-ray binaries, cataclysmic variables
etc).  Thus techniques based on well established methods of phase
sensitive detection were used to enhance the detection of $\gamma -$ rays
against the randomly arriving background \cite{pnb87,pnb90}.

This technique cannot be applied however for steady sources of $\gamma
-$rays like the Blazars, normal galaxies or unidentified $\gamma -$ray
sources discovered at $GeV$ energies.  There are, however, differences in
the detailed structure of the detected \v Cerenkov flash due to the fact
that the cosmic ray background protons are spatially and temporally
isotropic while the $\gamma -$ rays emanate from the point sources. In
addition, the physics of the hadronic cascades initiated by the protons in
the earth's atmosphere is different from the electromagnetic cascade
generated by the $\gamma-$ rays. Detailed monte carlo simulations have been
employed to study these differences. Techniques based on the shape of the
lateral distribution of the \v Cerenkov light pool to delineate the source
direction, have been studied \cite{ba93,on95} despite large
fluctuations in the measured \v Cerenkov photon densities \cite{vc98,vc99}. 
There are
several Atmospheric \v Cerenkov arrays designed precisely to apply these
techniques to ground based VHE $\gamma -$ ray astronomy 
\cite{pnb98,on96,cel96,arq97,oc97,tum90}. On the
other hand the imaging technique has been shown to be successful 
and also has been demonstrated to work reliably in detecting emission from
several $TeV$ $\gamma -$ ray sources \cite{hi85,we89}.

The spatial and temporal properties of the \v Cerenkov photons also contain
valuable information on the development and propagation of the EAS in the
atmosphere. As a result, systematic studies of these photons as received at
the observation level could lead to the development of techniques to
distinguish between hadronic or photon primaries. $\gamma -$ray primaries for 
example, develop higher in the atmosphere resulting in compact images as well
as narrower pulses. One would therefore expect correlations between different 
pulse shape parameters and imaging parameters.

Extensive studies have already been carried out in this regard using
detailed simulation techniques. Most of these studies were carried out at
higher energies with the aim of studying the elemental composition of
cosmic rays at these energies \cite{ch82}. The possibility of applying pulse 
shape
discrimination technique to improve the signal to noise ratio in the data from
Crab nebula was first demonstrated by T\"umer {\it et al.} \cite{tu90}. Their
study was solely based on the presence or absence of kinks and other
anomalies like long trailing edges of the \v Cerenkov pulses indicating
hadronic origin. These criteria were not based on any systematic
simulation studies nor were they conveniently parameterized so that it
could be used by others.

Several characteristics of the hadron- \& $\gamma -$ ray showers which were 
suggested  as possible discriminators in the past have been
met with limited success. Some of these are the presence of penetrating 
particles
among the EAS secondaries \cite{gr71}, the ultraviolet excess in hadron
initiated showers due to their proximity to the observation level \cite{st83}
and time duration \cite{fe68,re86}.  
On the other hand methods developed for improving the angular
resolution  technique \cite{gi82} are still in use and are 
one of the important methods of discrimination for  non-imaging arrays which
are generally spread out. 

A brief description of the efficacy of the pulse shape discrimination in
VHE $\gamma -$ ray astronomy was made by Patterson \& Hillas \cite{pat89}. Their
study was based solely on the presence of structure on the leading or
trailing edge of the \v Cerenkov pulses as suggested before \cite{tu90}. However
no systematic studies of the rise or decay times, FWHM etc were carried
out.  Recently Roberts {\it et al.} \cite{rb98}, developed a technique based
on the temporal \v Cerenkov pulse shape and showed that the use of rise time
at large zenith angles and FWHM at smaller zenith angles are effective
discriminators. However these studies are carried out at larger zenith
angle ($\geq 35^{\circ }$) and hence the conclusions are relevant only at
larger primary energies ($\geq 40~TeV$). On the other hand Cabot {\it et
al.} \cite{ca98} suggest possibility of identifying the muonic component, which
precedes the electromagnetic component, of the \v Cerenkov pulse in order
to separate the $\gamma-$ ray signal from the background. This could be a very
efficient technique in experiments based on wavefront sampling
technique. But once again this method works only at higher primary
energies ($\geq a\ few\ TeV$) where a significant number of muons are
produced.

In an earlier work\cite{vc99} functional fits have been carried out to the spherical
shower front to demonstrate that the radii of curvature are equal to the 
height of the shower maximum at all observation levels. It has also been 
demonstrated in that paper that the \v Cerenkov photon arrival time
distributions can be well fitted by a lognormal function. Using these fits, 
radial and spectral variation of pulse shape parameters have been
studied systematically.
 
In the present work we plan to make a systematic study of the temporal and
spatial profile of \v Cerenkov light from lower energy primaries both from
pure electromagnetic cascades as well as hadronic cascades generated by
$TeV$ energy primaries. The question we are trying to answer is whether
the observed differences in experimentally measurable temporal information
could be used to separate electromagnetic component from the hadronic
background.

In \S 2 of this paper details of simulations are given, followed by definition
of figure of merit for discrimination between $\gamma-$ rays and cosmic rays in
\S 3. In \S 4  we discuss the results from the analysis of the curvature of the 
\v Cerenkov front whereas in \S 5 we present results based on the pulse 
shape parameters and identify the more sensitive of them. In \S 6 we 
present the detailed study of the photon arrival time jitter. The dependence
of the parameters on the telescope opening angle, 
altitude of observation, incident angle of the primary as well as cosmic ray 
species and the shower core distance  
are discussed in \S 7. A brief discussion of the
results is presented in \S 8 and conclusions are summarized in \S 9.

\section{Simulations}

 A package called CORSIKA (version 560), \cite{kn98,hec98} has been used to simulate
\v Cerenkov light emission in the earth's atmosphere by the secondaries of
the extensive air showers generated by cosmic ray primaries or $\gamma-$  rays. 
This program simulates interactions of nuclei, hadrons, muons,
electrons and photons as well as decays of unstable secondaries in the
atmosphere. It uses EGS4 code \cite{ne85} for the electromagnetic
component of the air shower simulation and dual parton model for the
simulation of hadronic interactions at $TeV$ energies. The \v Cerenkov radiation 
produced
within the specified band width (300-650 $nm$) by the charged secondaries is
propagated to the ground.  The US standard atmosphere parameterized by
Linsley \cite{us62} has been used. The position, angle, time (with
respect to the first interaction) and production height of each photon
hitting the detector on the observation level are recorded.

       In the present studies we have mainly used Pachmarhi (longitude:
78$^{\circ}$ 26$^{\prime}$ E, latitude: 22$^{\circ}$ 28$^{\prime} N$ and
altitude: 1075 $m$) as the observation level where an array of \v Cerenkov
detectors each of area\footnote{This is the total reflective area 
of 7 parabolic mirrors of diameter 0.9 $m$ deployed paraxially on a single 
equatorial mount.} 4.35 $m^2$ is deployed in the form of a rectangular
array. We have assumed 17 detectors in the E-W direction with a separation of 25
$m$ and 21 detectors in the N-S direction with a separation of
20 m.  This configuration, similar to the Pachmarhi Array of \v Cerenkov
Telescopes (PACT) \cite{pnb98} but much larger, is chosen so that one can 
study the core
distance dependence of various observable parameters. 
Monoenergetic
primaries consisting of $\gamma-$ rays, protons and iron nuclei incident
vertically on the top of the atmosphere with their cores at the centre of
the array have been simulated in the present studies. The showers simulated in
this study have a fixed core position which is chosen to be the detector at the
centre of the array. The resulting \v Cerenkov pool is sampled by all the 357
detectors which are used to study the core distance dependence of the parameters
studied here. All the telescopes are assumed to have their optic axes aligned
vertically.

	An option of variable bunch size of the \v Cerenkov photons is
available in the package which serves to reduce the requirement of
hardware resources. However since we are interested in the fluctuations of
each of the estimated observables, we have tracked single photons for each
primary at all energies. Multiple scattering length for electrons and
positrons is decided by the parameter STEPFC in the EGS code which has
been set to 0.1 in the present studies \cite{fo78}. 
Wavelength dependent absorption of \v Cerenkov photons in the atmosphere
is not however taken into account.  The present conclusions
are expected to be independent of photon wavelengths. 

\section{Figure of merit of a parameter}

Figure of merit of a parameter that can distinguish between VHE $\gamma
-$rays and cosmic ray hadrons depends primarily on two factors.
Firstly, it should accept most of the $\gamma -$rays and secondly it
should be able to reject most of the hadrons. In general, this figure of
merit could be a function of primary energy. In the present work we define
such a figure of merit which is often called as {\it quality factor}, as
\cite{rb98}:

\begin{equation}
q={{N_a^{\gamma}} \over {N_T^{\gamma}}} \left( {{N_a^{cr}} \over {N_T^{cr}}} \right) ^{-{1 \over 2}}
\end{equation}
  
where $N_a^{\gamma}$ is the number of $\gamma$ rays accepted, 

$N_T^{\gamma}$ is the total number of $\gamma$ rays,

$N_a^{cr}$ is the number of background cosmic rays accepted and 

$N_T^{cr}$ is the total number of background cosmic rays.

The quality factor thus defined is independent of the actual number of 
$\gamma -$rays and protons recorded. 

In this paper, whenever a pair of distributions of a parameter under study for 
a hadron and a $\gamma -$ray primary are shown the threshold value  of the 
parameter is indicated by a vertical line. The threshold is chosen such that 
it yields maximum quality factor subject to the condition that the accepted 
fraction of $\gamma -$rays is $> 30\%$ and that of protons is $> 1\%$. The
errors on the qulaity factors shown in each case are statistical only.

\section{Shower front parameters}

It has been shown long ago that the radius of curvature of the \v Cerenkov
light front is strongly correlated with the height of shower maximum from
the observation level \cite{pr75}. This has been found
to be true for different species of cosmic rays \cite{vc99}. 
For photonic primaries the height of shower
maximum is decided by the radiation length in the atmosphere while that
for hadronic primaries it is decided by the interaction length which in turn
depends on the interaction cross-section in air.
Hence the radius of curvature could be species specific.
Therefore we have investigated the possibility of using the fitted radius
of curvature of the spherical \v Cerenkov front as a parameter to
distinguish between $\gamma -$ ray and proton initiated showers. The details of
a spherical fit to the mean arrival times of \v Cerenkov photons at detectors
sampling the front at various core distances are described in detail in \cite{vc99}. 
Fig. 1 shows the distribution of the fitted radii for different primary 
species of various energies. The primaries are incident vertically at the top 
of the atmosphere except wherever mentioned. Mono-energetic $\gamma -$ray and
hadron primaries of comparable \v Cerenkov yield are simulated and compared.
We also simulated showers, both for $\gamma -$
rays and protons whose energies are selected randomly from a power law
distribution of a differential slope of -2.65. This would also simulate the 
real data as one would record in an experiment. While the slope of the
$\gamma-$ ray spectrum could be different from what is chosen here, the 
quality factor does not depend on the number of showers and hence independent
of the spectral slope except when it shows strong energy dependence. The energy
bandwidth chosen here are 500 $GeV$ - 10 $TeV$ for
$\gamma-$ rays while it is 1 $TeV$ - 20 $TeV$ for protons.
 
Quality factors, as defined above, have been estimated for the present sample
from the distributions of the estimated radii of curvature for $\gamma -$ ray
and hadronic primaries as shown in figure 1. 
These  are given in table 1.
The threshold values in each case are 
indicated as vertical lines in figure 1.  Fraction of total number of
$\gamma -$ ray and proton showers that pass this cut, i.e., fraction of
showers with radii below the threshold as well as the number of showers 
simulated in each case are also tabulated.  The last row in the
table shows the quality factor estimated for primaries chosen from a powerlaw 
spectrum as mentioned before. This is consistent with that for mono-energetic 
primaries showing that a change in the height of shower maximum
with energy does not change the quality factor significantly. 
Fig. 1 shows the distributions of the fitted radii of curvature of the shower
front for proton primaries and $Fe$
primaries with respect to that for $\gamma -$ray primaries of equivalent 
\v Cerenkov yield. The vertical lines represent the threshold value with respect
to which quality factors are estimated.  One can see a rather high degree of 
overlap between the pair of distributions and hence the  quality factors
from this parameter are rather modest in value but are almost independent of  
primary energy.  However we will 
see in \S 7.1 that the use of a circular mask to limit the telescope
opening angle improves the quality factor at all
energies. Also the quality factor against $Fe$ primaries improves dramatically
suggesting that this could be the ideal parameter to discriminate heavy 
primaries. 
 
\begin{table}
\caption{Quality of radius of curvature of the spherical photon front as a 
discriminating parameter for vertical showers}
\vskip 0.3cm
\begin{tabular}{llllll}
\hline
\hline
Type of  & Energy of & Threshold & Fraction of  & Number of & Quality \\
primary  & primary   & radius of & showers  & showers  & factor \\
         & ($GeV$)     & curvature &   accepted  & simulated     &  \\
         &           &  ($km$)      &     (\%)    &     (\%)     & \\
\hline
$\gamma -$ rays & 100 & 11.7 &   99 & 200 & 1.13 $\pm$ 0.13 \\
 and protons    & 250 &      &  77.5  & 200  &       \\
\hline
$\gamma -$ rays & 500 & 8.8 & 98 & 200  & 1.13 $\pm$ 0.13 \\
 and protons    & 1000 &      &  75        & 200  &      \\
\hline
$\gamma -$ rays & 1000 & 8.2 &  99 & 100 & 1.11 $\pm$ 0.18\\
 and protons    & 2000 &     &  80    & 100  &      \\
\hline
$\gamma -$ rays & 1000 & 6.3 & 48   & 100  & 4.8 $\pm$ 2.6 \\
 and $Fe$ nuclei  & 10000 &     &  1       & 100  &      \\
\hline
$\gamma -$ rays & spectrum & 8.3 & 100 & 100  & 1.13 $\pm$ 0.18 \\
 and protons    & spectrum &     &  79      & 100  &      \\
\hline
\end{tabular}
\end{table}

\begin{figure}
\centerline{\psfig{file=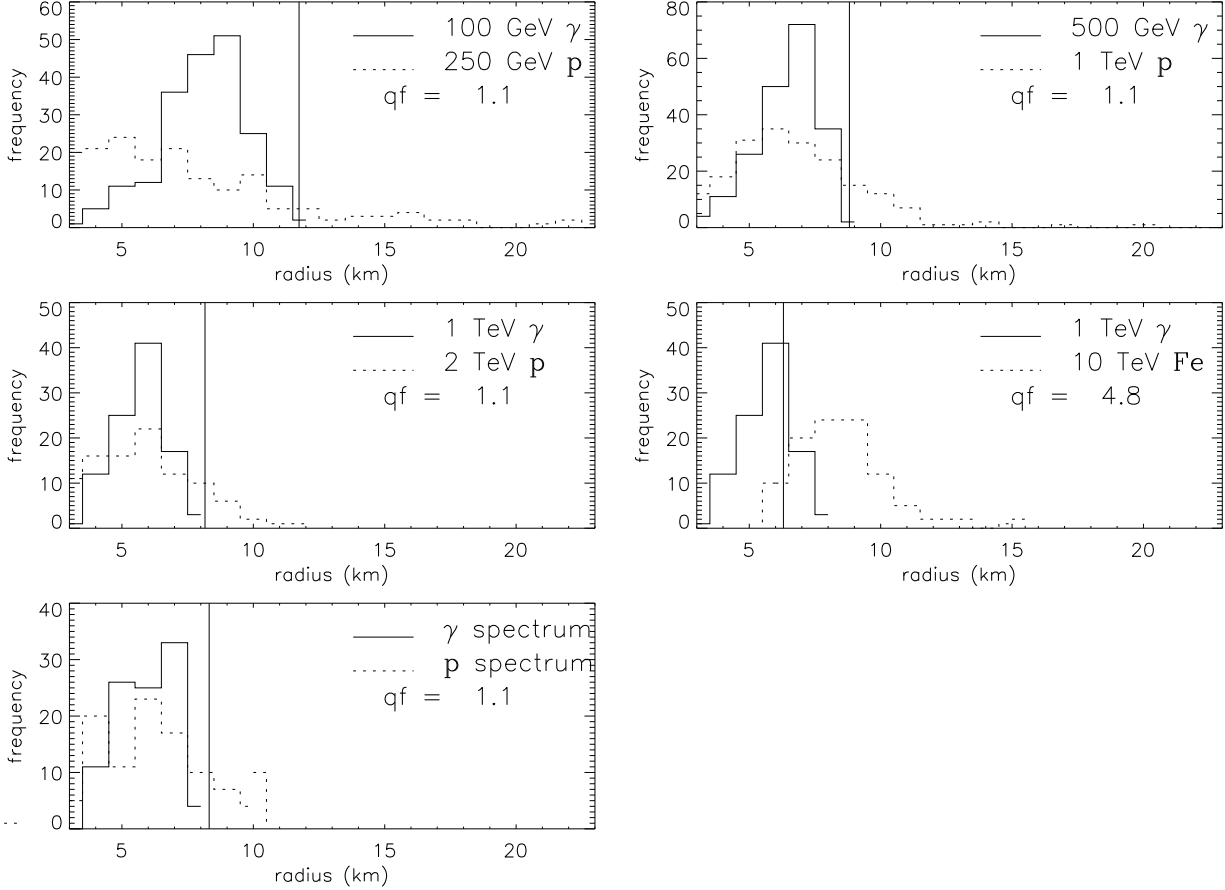,height=12cm }}
\caption{Distribution of the shower front radii for $\gamma -$ rays
(continuous line) and protons or $Fe$ nuclei (dashed line) of respective 
energies as indicated in each panel. The last panel shows similar distributions 
for $\gamma $-rays \& protons of primary energies chosen from a power law
spectrum of slope -2.65 (see text for details). The vertical lines indicate
the threshold values of parameters.}
\end{figure}

\section{Pulse shape parameters}

\subsection{Observation level of Pachmarhi}
Even though pulse shape parameters contain  information on the history of 
shower development
in the atmosphere, their use for identifying the primary species was always
in doubt especially at lower primary energies \cite{pat89}. On the other hand, 
it was 
known that the muons generated in hadron initiated showers will 
reach the
observation level several nanoseconds before the light from the electrons.
Consequently, the light from muons would give rise to an unmistakable precursor
which could be used as a discriminating parameter \cite{ca98}.  

Here we tried to investigate the possibility of using pulse shape parameters
to discriminate between $\gamma -$ ray and hadron showers, using the same data
set as was used for shower front paramemters. 
Distributions generated using pulse shape parameters were derived
from predicted lognormal distribution for individual detectors (see \cite{vc99} 
for details). It may be recalled here that the decay time of the pulse need not
be measured directly for this purpose. If one measures the mean arrival time and
the RMS fluctuations in \v Cerenkov photon arrival times at a detector, it will
enable us to derive the pulse shape parameters from the data directly \cite{vc99}. 
Among the three pulse shape parameters $viz.$ the rise time, the
decay time \& the pulse width, decay time seems to be the most sensitive 
parameter which exhibits species sensitive behaviour at $TeV$ energies.  However, 
\v Cerenkov photon arrival time distribution yields one set of pulse shape 
parameters for each of the 357 detectors. So also the arrival time jitter 
(discussed in \S 6). Hence for the purpose of estimating the statistical errors
on the quality factors, the sample size considered is the product of the number 
of detectors and the number of simulated showers. In order to take into account 
the core distance 
dependence of these parameters (see \cite{vc99} for details) they are averaged 
over 16\footnote{The choice of this number is purely arbitrary. We chose this
number because the spatial separation of the resulting 22 samples turned out 
to be reasonably uniform. However this grouping was not done or inclined showers.} consecutive detectors arranged in the order of 
increasing core distance.
The resulting 22 (352/16, excluding the detectors at the shower core and the 
farthest 4) sets of parameters are treated as independent samples
of the parameter under study. The distributions of the these samples of decay 
time both for
$\gamma -$ray and proton primaries of various energies  are shown in fig. 2.  
The number of showers simulated are same as those listed in table 1.
The distributions from hadronic primaries are in general characterized by 
longer tails compared to that from $\gamma -$ray primaries. There is a fair 
amount of overlap between the distributions of two types of primaries. 
All the relevant parameters including the fractions of $\gamma
-$rays and protons  whose decay times are less than the threshold value 
are listed in table 2.
In spite of the modest quality factor, one is able to reject
nearly 96\% of protons from the data while loosing nearly two thirds of the 
$\gamma -$ ray signal.

The other pulse shape parameters like the rise time and the pulse width
(defined as the full width at half maximum, FWHM) for vertical showers have 
been found to be quite
insensitive to the primary species and hence are not useful parameters to
distinguish hadronic events. However as discussed in \S 7.3 it can be seen 
that the pulse width could be a good parameter for inclined showers.

\begin{table}
\caption{Quality of pulse decay time as a discriminating parameter for vertical
showers. The number of showers simulated are same as those indicated in table 1.
}
\vskip 0.3cm
\begin{tabular}{lllll}
\hline
\hline
Type of  & Energy of & Threshold & Fraction of &  Quality \\
primary  & primary   &   value & showers  &  factor  \\
         & ($GeV$)     &   ($ns$)  & accepted  (\%)  &   \\
\hline
$\gamma -$ rays & 250 & 3.2 & 37.7 &  1.67 $\pm$0.02 \\
 and protons    & 500 &      & 5.1  &              \\
\hline
$\gamma -$ rays & 500 & 3.7 & 35   &       2.22 $\pm$ 0.03  \\
 and protons    & 1000 &      &  2.5 &             \\
\hline
$\gamma -$ rays & 1000 & 3.9 & 30.4 &       2.41 $\pm$ 0.06 \\
 and protons    & 2000 &     &  1.6 &             \\
\hline
$\gamma -$ rays & 1000 & 4.0 & 32.1 &        2.48 $\pm$ 0.06\\
and $Fe$ nuclei   & 10000 &    &  1.7 &             \\
\hline
$\gamma -$ rays & spectrum & 3.9 & 32.5 &        1.57 $\pm$ 0.03 \\
 and protons    & spectrum &     & 4.3  &             \\
\hline
\end{tabular}
\end{table}

\begin{figure}
\centerline{\psfig{file=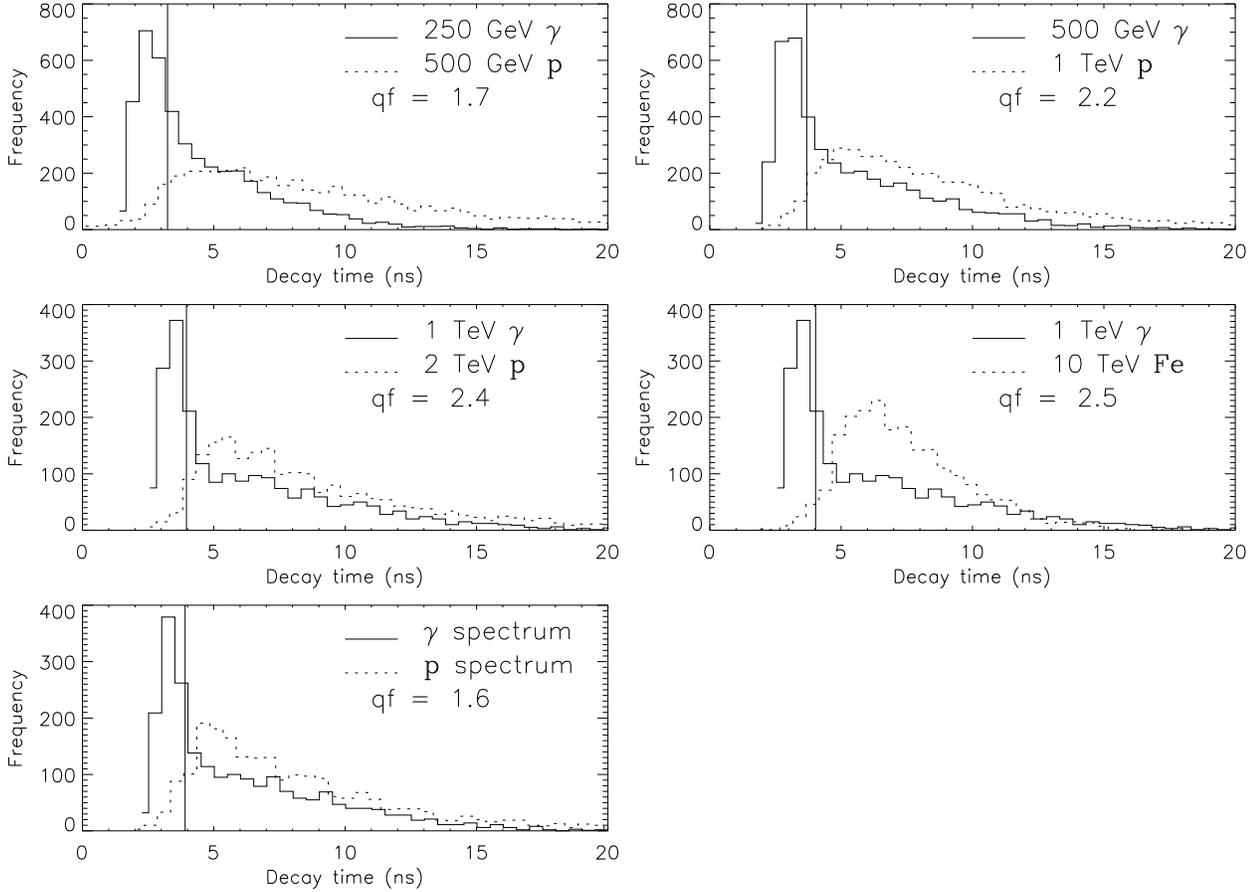,height=12cm}}
\caption{Distribution of decay times for $\gamma -$ rays (solid line) and 
protons or iron nuclei (dotted line). The vertical lines  indicate the threshold
values.}
\end{figure}

\section{Timing Jitter}

The \v Cerenkov light emitted at the top of the atmosphere reaches the observer
later than that from 
deeper down. The latter appears at larger angles but arrives earlier because the
particles travel faster than light in the atmosphere. It was 
suggested long ago that such small time differences could be useful in 
discriminating against proton showers \cite{pat89}. We have
seen before, that a majority of the \v Cerenkov photons originate at around
the shower maximum as a result of which at observation level we observe the 
spherical front
centred around that point. However the spread in the arrival time at any
core distance is largely decided by the photons emitted elsewhere in the
atmosphere. For example, at large core distances a bulk of the photons are
emitted at lower atmospheric heights, largely from lower energy electrons
undergoing multiple Coulomb scattering. In addition, photons seem
to arrive at increasingly larger angles away from the core thus implying 
larger arrival times. In other words, the spread in the arrival times
within a given shower at any core distance has the definite signature of
the kinematics of the shower development. Hence we tried to search for
species dependent signature in the arrival time spread of \v Cerenkov
photons. We quantify the arrival time spread in terms of the RMS of the
photon arrival times at a given detector. The ratio of this RMS to the mean
arrival time, called the relative jitter, is used as a discriminating 
parameter to identify the primary species. 

\subsection{Core distance dependence}
It may be recalled that each telescope in the array considered here consists of
compactly mounted 7 paraxial parabolic mirrors of diameter 0.9 $m$.
The average arrival time of \v Cerenkov photons is calculated
for each of the seven mirrors of a telescope, for a given shower. 
Then the mean arrival time for
each telescope is calculated which is the average of seven individual averages.
Time jitter is the RMS of seven averages for each telescope as mentioned
before. 
Figure 3 shows a radial  plot of the mean arrival times ($ns$, marked by +), time 
jitter ($ns$, marked by open triangles) and the relative
jitter (which is the ratio of time jitter to the mean, asterisks) both for
1 $TeV$ 
$\gamma -$rays (left) and 2 $TeV$ protons (right). While there is a general 
increase in
the jitter with increasing core distance for both the primaries, it shows
a minimum at around the hump region for $\gamma -$ray primaries. 
This is expected since most of the
photons here are emitted by the high energy electrons and also from a
limited range of atmospheric heights. Such a signature is swamped by kinematic
fluctuations in the case of hadronic primaries, in addition to exhibiting
an increased time spread \cite{vc98}.

\subsection{Separation of $\gamma$ - rays from proton primaries}

It may be seen from figure 3 that the relative time jitter for both $\gamma -$
ray as well as proton primaries is almost independent of
the core distance. Hence it is chosen as a  parameter for $\gamma -$hadron
separation even though it is not a necessary condition.  
  
\begin{table}
\caption{Quality of jitter as a discriminating parameter. The number of showers
simulated is 100 in all cases for each type of primary.
}
\vskip 0.3cm
\begin{tabular}{lllll}
\hline
\hline
Type of  & Energy of & Threshold & Fraction of &  Quality \\
primary  & primary   &   value   & showers  &  factor  \\
         & ($GeV$)     &       &   accepted (\%)   &   \\
\hline
$\gamma -$ rays & 100 & 0.23 & 49.8 & 2.83 $\pm$ 0.05  \\
 and protons    & 250 &     & 3.1 &       \\
\hline
$\gamma -$ rays & 1000 & 0.07 & 67.5 & 2.42 $\pm$ 0.03 \\
 and protons    & 2000 &     &  7.8   &       \\
\hline
$\gamma -$ rays & 1000  & 0.09  & 91.3 & 9.13$\pm$ 0.25 \\
 and $Fe$ nuclei  & 10000 &     & 1.0 &         \\
\hline
$\gamma -$ rays & spectrum & 0.08 & 80.5 &   1.85$\pm$ 0.02 \\
 and protons    & spectrum &     & 19   &        \\
\hline
\end{tabular}
\end{table}

\begin{figure}
\centerline{\psfig{file=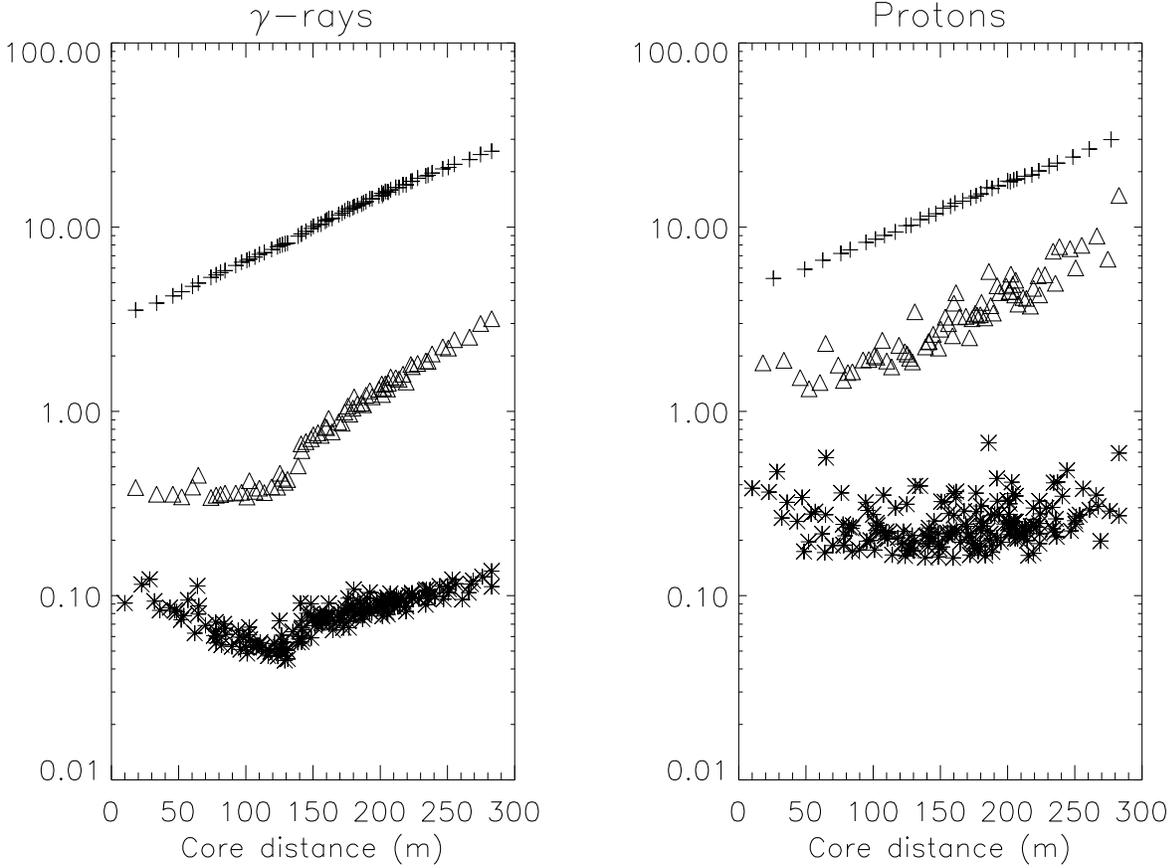,height=12cm}}
\caption{Radial variation of the mean arrival time, $ns$ (+), arrival time 
jitter, $ns$ (open triangle)
and the relative jitter (asterisk) for 1 $TeV$ $\gamma -$ray (left) and 
2 $TeV$ proton (right) incident vertically at the top of the atmosphere.}
\end{figure}

Fig. 4 shows the comparative distributions of this parameter for $\gamma -$rays
and protons (a \& b) as well as $\gamma -$rays and  $Fe$ primaries (c) of 
equivalent \v Cerenkov yields.  Also shown in the figure (d) are comparative 
distributions for $\gamma -$ray and 
proton primaries of varying energies selected from a power law distribution of 
slope  -2.65. The energy bandwidths chosen for $\gamma -$rays and protons are
0.5 - 10 $TeV$ and 1 - 20 $TeV$ respectively. The results are based on 100 
showers in each case for each species.

It is well known that after shower maximum, $\gamma -$ray showers attenuate 
progressively faster with atmospheric depth than do hadronic showers 
\cite{pat83}.
In addition, the lateral spread also is larger for hadron
initiated showers. As a result, the RMS spread in the time of arrival of 
photons at the observation level is expected to be larger for hadronic
showers. Consequently, the
distributions in fig 4 are pretty wide with a long tail for hadrons at all 
primary energies which makes it possible to discriminate them.

Table 3 shows the quality factors derived from relative jitter for vertical
showers of $\gamma -$rays, protons and iron primaries.  
From the table it can be seen that using relative
jitter as a discriminating parameter one can reject more than $\sim $80\% of protons
while retaining more than $\sim $80\% of $\gamma -$rays (from spectrum).

\begin{figure}
\centerline{\psfig{file=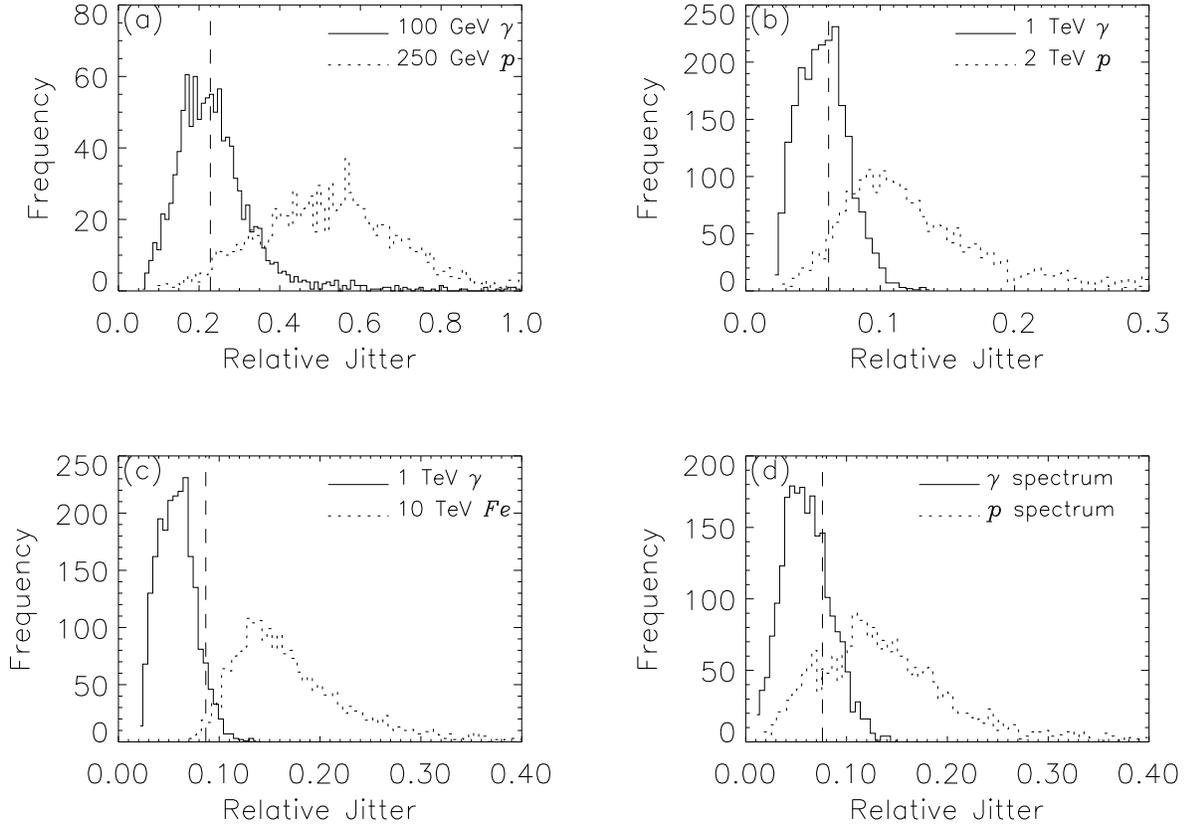,height=12cm}}
\caption{Distributions of mean relative jitter for $\gamma -$rays (solid line)
and protons or iron nuclei (dotted line) of different primary energies are 
shown. (d) shows the
distribution for primaries with a power law (differential slope -2.65) 
distribution of 
energies in the range 500 $GeV$ - 10 $TeV$ for $\gamma -$rays and 1 $TeV$ - 20 $TeV$ for
protons. The vertical lines represent the threshold values.
}
\end{figure}

\section{Sensitivity to various operational parameters}
 
\subsection{Effect of telescope opening angle on quality factors}

The opening angle of a \v Cerenkov telescope is often limited by placing a 
circular mask 
at the focal point in front of the photocathode.  This limits the arrival
angle of the photons reaching the photocathode. In the absence of a mask the
opening angle is limited by the photocathode diameter. In other words, the 
limiting
mask is expected to reduce the mean arrival angle of photons, reduce the mean
arrival time as well as increase the average production height. Table 4
summarizes the effect on aforesaid shower parameters by  placing a $3^\circ$ 
(FWHM) mask at the focal point of a detector. This is to demonstrate 
qualitatively its effect on some of the parameters of shower development. 
Vertically incident showers are 
considered here. The changes in the mean angle of arrival and time  is around
30\% for $\gamma -$rays while it is significantly larger for proton primaries.
The mean production height is much less sensitive to the presence of mask which
increases the mean production height by about 10\% and 14\% for $\gamma -$ray
and proton primaries respectively.

The question we are addressing here is the effect of telescope opening angle on
each of the quality factors discussed here. Tables 5, 6, 7 \& 8 summarize the 
results relating to the role of mask when radius of curvature, pulse decay time,
pulse width and relative jitter are used as parameters to discriminate against 
cosmic ray hadrons. The same data set is used to compare the results with and
without mask.  In the no-mask case all the \v Cerenkov photons incident on a
detector, irrespective of their angle of incidence, are accepted. Pulse width 
was included in this case to show that if a mask is
in use it could be a useful parameter to discriminate against heavy primaries. 
Its sensitivity improves with increasing primary energy consistent with
other such studies.
From tables it can be
readily seen that the use of a mask ($5^\circ$ FWHM) improves the 
efficiency of the radius of curvature as a
discriminating parameter while the pulse shape parameters show a marked 
improvement only for heavy primaries.  For the relative timing
jitter the efficiency as a discriminating parameter reduces when a mask is
used. This, even though a marginal effect, is quite understandable.  
Limiting the photon arrival angles to
near vertical direction reduces the RMS fluctuations because of limited range
of pathlength differences.

\begin{table}
\caption{Effect of adding a mask at the focal point on some of the parameters
for vertical showers}
\vskip 0.3cm
\begin{tabular}{llllll}
\hline
\hline
Type of  & Energy of & Mask & Mean  & Mean  & Mean \\
primary  & primary   & used & arrival angle  & arrival time & production   \\
         & ($GeV$)     & (FWHM) & (degrees)        & ($ns$)      & height ($km$) \\

\hline
$\gamma -$ray & 100 & No mask & $1.3 \pm 0.27$ & $7.23 \pm 1.91$ & $11.04 \pm 1.15$ \\
\hline
$\gamma -$ray & 100 & $3.0^\circ $ & $0.9 \pm 0.11$ & $5.11 \pm 0.66$ & $12.09 \pm 1.13  $\\
\hline
Proton  & 250 & No mask & $1.64 \pm 0.42$ & $12.01 \pm 18.29$ & $7.83 \pm 1.
11$ \\
\hline
Proton  & 250 & $3.0^\circ $ & $0.93 \pm 0.11$ & $7.11 \pm 14.41$ & $8.91 \pm 1.11 $\\
\hline
\end{tabular}
\end{table}

\begin{table}
\caption{Quality of radius of spherical wavefront as a discriminating parameter
for vertical showers when $5^\circ$ mask is used.}
\vskip 0.3cm
\begin{tabular}{llllll}
\hline
\hline
Type of  & Energy of & Threshold & Fraction of &  Quality \\
primary  & primary   & radius of & showers  &  factor  \\
         & ($GeV$)     & curvature & accepted     &   \\
         &           &  ($km$)      &   (\%)       & \\
\hline
$\gamma -$ rays & 100 & 12.5  & 97 &  1.37 $\pm$ 0.16 \\
 and protons    & 250 &      &  50.5  &        \\
\hline
$\gamma -$ rays & 500 & 9.7 & 96.5 &  1.35$\pm$ 0.16 \\
 and protons    & 1000 &      &  51   &      \\
\hline
$\gamma -$ rays & 1000 & 8.8 & 93 &  1.33 $\pm$ 0.22 \\
 and protons    & 2000 &     & 49   &       \\
\hline
$\gamma -$ rays & 1000 & 8.7 &  91 &  9.1 $\pm$ 4.8 \\
 and $Fe$ nuclei  & 10000 &     & 1.0        \\
\hline
$\gamma -$ rays & spectrum & 9.5 & 100 &  1.3  $\pm$ 0.2\\
 and protons    & spectrum &     & 61    &      \\
\hline
\end{tabular}
\end{table}

\begin{table}
\caption{Quality of decay time as a discriminating parameter with $5^\circ$ 
mask}
\vskip 0.3cm
\begin{tabular}{llllll}
\hline
\hline
Type of  & Energy of & Threshold & Fraction of &  Quality \\
primary  & primary   &   value & showers        &  factor  \\
         & ($GeV$)     &   ($ns$)  & accepted (\%)    &    \\
\hline
$\gamma -$ rays & 250 & 3.4 & 52.8   &        1.14$\pm$ 0.01 \\
 and protons    & 500 &      & 21.5   &              \\
\hline
$\gamma -$ rays & 500 & 3.5 & 44.9   &       1.84 $\pm$ 0.02 \\
 and protons    & 1000 &      &  6    &             \\
\hline
$\gamma -$ rays & 1000 & 3.5 & 40.1   &        1.81$\pm$0.03 \\
 and protons    & 2000 &     & 4.9  &             \\
\hline
$\gamma -$ rays & 1000 & 4.8 & 71.4 &          6.98 $\pm$ 0.19 \\
and $Fe$ nuclei   & 10000 &    & 1.1  &             \\
\hline
$\gamma -$ rays & spectrum & 3.3 & 35.2 &       1.70 $\pm$0.03 \\
 and protons    & spectrum &     &     4.3 &             \\
\hline
\end{tabular}
\end{table}

\begin{table}
\caption{Quality of FWHM as a discriminating parameter when $5^\circ$ mask is
used.}
\vskip 0.3cm
\begin{tabular}{lllll}
\hline
\hline
Type of  & Energy of & Threshold & Fraction of  &  Quality \\
primary  & primary   &   value & showers          &  factor  \\
         & ($GeV$)     &   ($ns$)  &  accepted (\%)   &    \\
\hline
$\gamma -$ rays & 250 & 3.8 & 98.2   &        1.02$\pm$0.01 \\
 and protons    & 500 &      & 92.5   &              \\
\hline
$\gamma -$ rays & 500 & 2.0 & 45.7   &        1.08$\pm$0.01  \\
 and protons    & 1000 &      & 18.1   &             \\
\hline
$\gamma -$ rays & 1000 & 2.1 & 39.3   &        1.32$\pm$0.02 \\
 and protons    & 2000 &     & 8.9    &             \\
\hline
$\gamma -$ rays & 1000 & 2.0 & 33.8   &        3.31 $\pm$0.09 \\
and $Fe$ nuclei   & 10000 &     & 1.1    &            \\
\hline
$\gamma -$ rays & spectrum & 2.2 & 45   &        1.07 $\pm$0.01 \\
 and protons    & spectrum &     &  17.6  &             \\
\hline
\end{tabular}
\end{table}

\begin{table}
\caption{Quality of jitter as a discriminating parameter when a $5^\circ$ mask 
is used.}
\vskip 0.3cm
\begin{tabular}{lllll}
\hline
\hline
Type of  & Energy of & Threshold & Fraction of  &  Quality \\
primary  & primary   &   value & showers          &  factor  \\
         & ($GeV$)     &     & accepted (\%)    &   \\
\hline
$\gamma -$ rays & 100 & 0.2  & 58.2 &       1.7  $\pm$ 0.02 \\
 and protons    & 250 &     & 12.3 &              \\
\hline
$\gamma -$ rays & 1000 & 0.04  & 62.1 &        2.1 $\pm$ 0.03 \\
 and protons    & 2000 &     &   9.1  &              \\
\hline
$\gamma -$ rays & 1000  & 0.05  & 89.6 &        8.8 $\pm$ 0.2  \\
 and $Fe$ nuclei  & 10000 &     &  1.1  &              \\
\hline
$\gamma -$ rays & spectrum  & 0.05  & 86.4 &          1.7 $\pm$ 0.02 \\
 and protons    & spectrum  &     &  26.7 &              \\
\hline
\end{tabular}
\end{table}

\subsection{Altitude dependence of quality factors}

As the altitude of observation increases, the shower maximum for a given 
primary energy comes closer to the observation level. As a result, 
the lateral distribution of \v Cerenkov photons has been known to change 
with the altitude of the observation level. The core distance at which the
hump appears as well as the prominence of the hump will be smaller with
increasing altitude of observation \cite{ra88,po98}. Since the \v Cerenkov front
is being intercepted at different observation levels during its propagation in 
the atmosphere, the shower parameters like the 
average arrival angle, time $etc$ also will be different at different 
observation levels. As we will see in \S 7.5  that the species discriminating 
efficiency of
the parameter does vary, even though weakly, with core distance. Hence it 
would be interesting to see if their role depends on the observation level too.
Therefore we studied the role played by the observation altitude in using 
the various types of parameters studied here.
  
\begin{table}
\caption{Quality of decay time as a parameter measured at sea level when
used to discriminate 500 $GeV$ $\gamma -$rays from 1 $TeV$ protons.}
\vskip 0.3cm
\begin{tabular}{lllll}
\hline
\hline
Type of  & Mask & Threshold & Fraction of  &  Quality \\
primary  & used ($^\circ$) &   value & showers          &  factor  \\
         &  (FWHM)     &   ($ns$)  &  accepted (\%)   &   \\
\hline
$\gamma -$ rays & No mask & 3.2 & 32.9   &       2.82 $\pm$ 0.07 \\
 and protons    & used &      & 1.4      &             \\  
\hline
$\gamma -$ rays & $5^\circ$ mask & 3.0 & 37.5   &       3.45$\pm$0.09 \\
 and protons    & used &      & 1.2    &              \\
\hline

\end{tabular}
\end{table}

\begin{table}
\caption{Quality of pulse decay time as measured at an altitude of
2.2 $km$ when used to discriminate 500 $GeV$ $\gamma -$rays from 1 $TeV$ protons.}
\vskip 0.3cm
\begin{tabular}{lllll}
\hline
\hline
Type of  & Mask  & Threshold & Fraction of &  Quality \\
primary  & used ($^\circ$)  &   value & showers         &  factor  \\
         & (FWHM) &   ($ns$)  &  accepted (\%)  &   \\
\hline
$\gamma -$ rays & No mask & 4.2  & 31.5   &        1.07$\pm$0.02  \\
 and protons    & used  &      & 8.6    &             \\  
\hline
$\gamma -$ rays & $5^\circ$ mask & 6.8 & 97.3   &        1.06 $\pm$0.01  \\
 and protons    & used &      & 84.1   &             \\

\hline
\end{tabular}
\end{table}

\begin{table}
\caption{Quality of jitter as a discriminating parameter at two observation 
levels $viz$ sea level \& 2.2 $km$ above mean sea level. Monoenergetic 
$\gamma -$rays of 500 $GeV$ and protons of
1 $TeV$ incident vertically at the top of the atmosphere are considered.}
\vskip 0.3cm
\begin{tabular}{llllll}
\hline
\hline
Observation & Mask  & Threshold & Fraction of & Fraction of &  Quality \\
level above  & used ($^\circ$)  &   value &  accepted  & accepted &  factor  \\
msl ($km$)         & (FWHM)     &     & $\gamma -$rays(\%)  &  protons(\%)   &  \\
 & & & & ($\gamma$, p) &  \\
\hline
0.0  & None & 0.08 & 52.5 &  2.7 &         3.18 $\pm$ 0.06  \\
\hline
1.07 & None & 0.09 & 45.8 &  1.1 &         4.48 $\pm$ 0.12  \\
\hline
2.2  & None & 0.09 & 41.2 &  1.7 &         3.16 $\pm$ 0.07 \\
\hline
0.0  & 5.0  & 0.05 & 41.0 &  3.3 &  2.27 $\pm$ 0.03 \\
\hline
1.07 & 5.0  & 0.05 & 48.9 &  3.4 &  2.67 $\pm$ 0.05 \\
\hline
2.2 &  5.0  & 0.05 & 57.3 &  4.5 &  2.7 $\pm$ 0.04 \\
\hline
\end{tabular}
\end{table}

Tables 9, 10 \& 11 summarize the quality factors for the pulse decay times as
well as the relative timing jitter at two different observation levels $viz.$
sea level and 2.2 $km$ above mean sea level (for 500 $GeV$ $\gamma -$rays and 
1 $TeV$ protons). In comparison with the quality
factors for the same parameters in table 2 \& 3 it can be seen that  the
quality factors from pulse decay time improve steadily with decreasing altitude
while those from timing jitter are almost independent of altitude. In order to 
understand this we computed the quality factors, for detectors only around the 
hump region after taking into account the varying hump distances from the core 
at three altitudes. It is found that the quality factors so calculated for pulse
decay time are respectively 8.1, 6.6, and 3.6 for sea level, 1 $km$ and 
2.2 $km$ altitudes.  As one would see later (table 16) that the quality factors 
computed exclusively for pre-hump \& post-hump regions are poorer than that for 
the hump region.  
Hence the improvement in quality factor at lower altitudes  is 
mainly due to the increased prominence of hump at lower atmospheric depths.

The relative timing jitter, on the other hand seems to be much less sensitive 
to core distance (see figure 3) and hence doesn't vary significantly with 
observation altitude.

\subsection{Incident angle dependence of quality factors}

At larger zenith angles rising absorption and the increasing distance from the
shower maximum raise the energy threshold of the primary that are detected by
an atmospheric \v Cerenkov telescope. After the shower maximum $\gamma -$ray 
induced showers attenuate progressively faster with atmospheric depth than do 
the hadronic showers. As a result one would expect a zenith angle dependence
on the sensitivity of the parameters studied here. 
Qualitatively speaking inclined showers at a given altitude behave similar to
vertical showers at a lower altitude. As a result, one would expect the 
quality factor to improve with zenith angle, as it would with a decrease in the 
altitude of observation. 
The species sensitive imaging parameters like the azwidth for example, have been
shown to be much less sensitive for primaries incident at angles $\ge 30^\circ$ 
\cite{we89}. Hence we wanted to check the possible dependence  of the 
parameters, discussed here,  on zenith angle of the primary. We have
generated showers initiated by $\gamma -$rays and hadrons incident at the top 
of atmosphere at a fixed zenith angle of $30^\circ$ to the vertical and 
uniformly distrbuted in azimuth. 

The quality factors have been derived at one test energy for comparison for
all the parameters under consideration in the present study. The
results are summarized in the following tables.

\begin{table}
\caption{Quality of radius of spherical wavefront as a discriminating parameter
for inclined showers (incident at  $30^\circ$ to the vertical).}
\vskip 0.3cm
\begin{tabular}{llllll}
\hline
\hline
Type of  & Energy of & Threshold & Fraction of  & Fraction of & Quality \\
primary  & primary   & radius of & $\gamma -$ rays  & protons  & factor  \\
         & ($GeV$)     & curvature &   accepted     &   accepted       &  \\
         &           &  ($km$)      &    (\%)      &  (\%)            & \\
\hline
$\gamma -$ rays & 500 & 10.9 & 94 & 51   & 1.32 $\pm$  0.16 \\
 and protons    & 1000 &      &      &      &      \\
(without mask) &&&&&\\
\hline
$\gamma -$ rays & 500 & 11.6 & 93.5 & 46.5 & 1.37 $\pm$ 0.16 \\
 and protons    & 1000 &     &      &      &      \\
(with $5^\circ$ mask) &&&&&\\
\hline
\end{tabular}
\end{table}

Table 12 summarizes the efficiency of the radius of curvature of the light
front as a discriminant for inclined showers of energy 500 $GeV$ and 1 $TeV$
for $\gamma -$ray and proton primaries respectively. The results are based on 
200 showers each of $\gamma -$rays and protons.  It can be readily seen from
a comparison with table 1 that there is a significant improvement in the quality
factors for inclined showers.  This could be partly due to an increase in the 
range of fitted radii of proton showers. 

\begin{table}
\caption{Quality from decay time for showers incident at $30^\circ$ to the 
vertical when used to discriminate 500 $GeV$ $\gamma -$rays from 1 $TeV$ protons.
}
\vskip 0.3cm
\begin{tabular}{llllll}
\hline
\hline
Type of  & Mask dia. & Threshold & Fraction of & Fraction of & Quality \\
primary  & used   &   value &  $\gamma -$ rays  & protons  & factor  \\
         & (FWHM)     &   ($ns$)  & accepted (\%)  &  accepted (\%)   &  \\
\hline
$\gamma -$ rays & No mask  & 5.4 & 59.9   & 11.3   & 1.79 $\pm$ 0.02 \\
 and protons    & used  &     &      &      &       \\
\hline
$\gamma -$ rays &  $5.0^\circ$   & 4.6 & 52.5   &  8.2   & 1.83 $\pm$ 0.02 \\
 and protons    &   &     &      &      &       \\
\hline
\end{tabular}
\end{table}

\begin{table}
\caption{Quality of pulse width (FWHM) as a parameter from showers incident at
$30^\circ $ to the vertical
when used to discriminate 500 $GeV$ $\gamma -$rays from 1 $TeV$ protons.}
\vskip 0.3cm
\begin{tabular}{llllll}
\hline
\hline
Type of  & Mask dia. & Threshold & Fraction of & Fraction of & Quality \\
primary  & used   &   value & $\gamma -$ rays  & protons  & factor  \\
         & (FWHM)     &   ($ns$)  & accepted  (\%)  & accepted  (\%)  &  \\
\hline
$\gamma -$ rays & No mask & 3.9 & 84.3   & 48.8   & 1.21$\pm$ 0.01  \\
 and protons    & used &      &      &      &      \\
\hline
$\gamma -$ rays & $5.0^\circ$  & 3.4 & 75.8   & 33.6   & 1.31 $\pm$ 0.01  \\
 and protons    &      &     &      &      &      \\
\hline
\end{tabular}
\end{table}

	Tables 13 \& 14 summarize the results of using two of the pulse shape
parameters $viz.$ pulse decay time and pulse width for species identification
for showers incident at $30^\circ$ to the vertical. The results are based on
100 showers each of $\gamma -$rays \& protons. Keeping in mind that the 
decay times are not grouped for inclined showers (unlike vertical showers)
quality factor for decay time improves for inclined showers.

	Table 14 on the other hand has an interesting result. As mentioned
in \S 5.1, pulse rise time and pulse width do not exhibit significant 
sensitivity to primary species for vertical showers. However for inclined
showers pulse width becomes a useful parameter as shown in table 14. 
This is consistent with the results of Roberts $et~al.$ \cite{rb98}, who 
showed that the 
quality factor for pulse width peaks at around an inclination of $\sim 35^\circ$
to the vertical at a given altitude of observation.  

 The relative shower to shower fluctuations in pulse shape parameters 
have been found to be less for inclined showers \cite{vc99} unlike at higher
energies where the trend reverses \cite{rb98}. Table 15 summarizes 
the quality factors using the timing  jitter as the parameter. 
It may be seen from a comparison of tables 3 \& 15 that the species sensitivity 
of the timing jitter falls with zenith angle of the primary. 

The sensitivity of each of the parameters for inclined showers has been 
further studied after introducing a focal point mask. The results are again
summarized in tables 12-15. The opening angles of inclined showers are
smaller than that for vertical showers because, inclined 
showers reach shower maximum at a higher altitude where the \v Cerenkov angle is
smaller. Further, the lower energy electrons propagate in a lower density
medium thus undergoing less frequent Coulomb scattering. Consequently, the
\v Cerenkov emission from inclined showers is more collimated compared to that
from a same energy primary in vertical direction. Hence the use of a focal 
point mask is not expected to
change the quality factors significantly since a smaller
fraction of photons are obstructed by it.
The results in tables 12-15 show no major change in quality factors when focal
point masks are introduced, essentially supporting the above argument.

For reasons mentioned above, the differential pathlengths of \v Cerenkov 
photons also reduce resulting in a poor sensitivity for timing jitter for
inclined showers compared to that for vertical showers. The results of table 15
support this conclusion. Once again the use of a focal point mask has no
significant effect on the quality factor. However at higher primary energies
it is observed that the spread in the arrival times of \v Cerenkov photons 
increases with the mass of the primary and the zenith angle \cite{rb98}. This 
is due to the increased muon production in cascades generated by the heavier
primaries.

\begin{table}
\caption{Quality of relative jitter as a discriminating parameter
for $\gamma -$ray (500 $GeV$) and proton (1 $TeV$) primaries incident at $30^\circ$
with respect to the vertical.}
\vskip 0.3cm
\begin{tabular}{llllll}
\hline
\hline
Type of  & Mask dia. & Threshold & Fraction of & Fraction of & Quality \\
primary  & used   &   value & $\gamma -$ rays & protons & factor  \\
         & (if any)     &     &  accepted (\%)  &  accepted (\%) &  \\
\hline
$\gamma -$ rays & No mask  & 0.006  & 66.3 & 37.5  & 1.08$\pm$0.01 \\
 and protons    & used &     &      &      &       \\
      &      &&&&\\
\hline
$\gamma -$ rays & $5.0^\circ$  & 0.009 & 99.8 & 98.7  & 1.01$\pm$0.01 \\
 and protons    &  &     &      &      &       \\
      &      &&&&\\
\hline
\end{tabular}
\end{table}

\subsection{Species dependence of quality factors}

Each of the three species sensitive parameters under study here were applied to
$Fe$ primaries as well. It can be easily seen that in all the cases these
parameters are more efficient in discriminating heavy primaries as compared to
protons. The radius of the shower front, for example, shows a higher 
sensitivity for $Fe$ primaries which is a reflection of the fact that,  
an $Fe$ shower reaches its maximum development at a higher
altitude compared to that of a lower $Z$ primary. The efficiency also improves 
significantly with the use of a mask, a feature similar to that of proton 
primaries.

The pulse shape parameters on the other hand do not exhibit a large change in 
sensitivity for high $Z$ primaries. However the use of mask shows a drastic 
improvement in sensitivity of decay time as the quality factor changes from 
$\sim 2.5$ (with out mask) to $\sim 7$ (with a $5^\circ$ mask). 

The photon arrival time jitter is particularly sensitive to high $Z$ primaries.
This perhaps is quite understandable since the $Fe$ shower if considered as 
superposition of proton showers is expected to be more spread out
laterally and hence the timing jitter will deviate more than that 
for $\gamma -$ray showers.

This demonstrates that the parameters discussed here are sensitive not only to
proton primaries but also to high $Z$ primaries, unlike Hillas's image 
parameters which are not very sensitive to heavy primaries of cosmic rays.

\subsection{Core distance dependence of quality factors}

\begin{table}
\caption{Quality of decay time as a discriminating parameter
for $\gamma -$ray (500 $GeV$) and proton (1 $TeV$) primaries at
three different core distance ranges: pre-hump, hump \& post hump}
\vskip 0.3cm
\begin{tabular}{llllll}
\hline
\hline
Core distance  & Mask dia. & Threshold & Fraction of & Fraction of & Quality \\
ranges  & used   &   value & $\gamma -$ rays & protons & factor  \\
         & (if any)     &   ($ns$)  &  accepted (\%)   & accepted (\%)    &  \\
\hline
Pre-hump & No Mask & 4.3 & 88.5 & 30.5 & 1.60 $\pm $ 0.05 \\
\hline
Hump     & No Mask & 3.3 & 80.5 &  1.5 & 6.57 $\pm$ 0.51 \\
\hline
Post-hump & No Mask &   8.4 & 53.0 & 14.0 & 1.42 $\pm$ 0.06 \\
\hline
Pre-hump  & $5.0^\circ$ & 3.8 & 73.0  & 19.5   & 1.65 $\pm$ 0.06 \\
\hline
Hump      & $5.0^\circ$ & 2.8 & 91.5  &  1.0   & 9.15 $\pm$ 0.85 \\
\hline
Post-hump & $5.0^\circ$ & 6.0 &  93.5 & 41.5 & 1.45 $\pm$ 0.04 \\
\hline
\end{tabular}
\end{table}

There are mainly three effects that determine the lateral distribution of 
\v Cerenkov light at the observation level: the finite \v Cerenkov angle,
Coulomb scattering of the  progenitor electrons and the transverse momentum $p_t$, 
in hadronic interactions. While the first effect is mainly responsible for the
proverbial `hump' in the lateral distribution of $\gamma -$ray initiated 
showers the third effect is responsible for the observable differences between
the two types of primaries and the second effect is responsible for smoothing
out the other two effects. As a result, one expects species specific parameters to
show a dependence on the core distance of the detectors which measure these
parameters. Table 16 summarizes the relative efficiencies of one of the  
parameters, $viz$ pulse decay time as measured by detectors at pre-hump, hump 
\& post-hump regions of showers. It can be seen that the difference between 
$\gamma -$ray and proton initiated showers is maximum at around the hump
region resulting in maximum quality factor at a given primary energy. At the
other two regions, $viz.$ pre-hump \& post-hump regions, the quality factors 
are comparable. However in
practice it may not be always possible to detect the hump. Hence it may be
advisable to limit the core distance of the triggering events to within the 
hump region at a given altitude of observation which is quite feasible since
hump radius is not very sensitive to primary energy. 

\section{Discussions}

In the present study we did not take into account the effect of detector 
response on the parameters selected here. It is implicit that in an experiment
the pulse shapes have to be measured using high bandwidth electronics to
minimize the shape distortion due to instrumental response so that the pulse 
shape parameters could be effective discriminators.  However while 
applying these techniques to real data, it is inevitable that the theoretical
parameters discussed here be suitably corrected for instrumental response used
in the system. The threshold values of the parameters have to be derived for
a given detection system and then applied to the data in order to reject events 
of hadronic origin. 

The quality factors derived from the current studies have
been tested for stability by estimating the same for subsets of the totality of
data used. Hence their values are not critically dependent on the sample 
size. However it must be mentioned we chose to use that value of the parameter
averaged over 16 detectors, thus reducing the 357 radial samples to 22. 
Obviously, the quality factors will improve if one averages over a larger number
of detectors, as has been seen in our analysis. This only means that arrays
with larger number of detectors can reject hadronic showers with a better
efficiency. 

The effect of mask has been studied qualitatively here. In order to apply the 
technique to real data one has to use the actual mask diameter. 
It is possible, in principle, to optimize the mask diameter for the best hadron
rejection efficiency.

\subsection{Radius of curvature of the shower front:} 

From the table 1 it 
can be seen that the radius of curvature of the \v Cerenkov front  can enable 
us to reject around 20\% of proton
events at all energies. The rejection efficiency could improve to 50\% if one
uses a mask to limit the photon incidence angle. On the other hand this 
parameter is very sensitive to 
heavy
primaries and hence  one can reject up to $\sim 99\%$ (with or without a 
$5^\circ $ mask) of them. 

At higher primary energies ($>~20$ $TeV$), the height of shower maximum has 
been shown to be directly proportional to the slope of the \v Cerenkov 
photon lateral distribution \cite{arq96}. This has been 
suggested to be a good parameter to distinguish between $\gamma -$ rays and 
hadrons. The present results show that the height of shower maximum, 
measured as the fitted curvature of the light front, is a
reasonably good parameter even at lower energies and the sensitivity improves
for inclined showers. As already mentioned, it is particularly useful to 
discriminate  against heavy 
primaries. It can be readily seen from tables 1 \& 5 that the quality factor is
sensitive to photons arriving at large incidence angles. This is particularly so
for heavy primaries.

However the conventional method of measuring the height of shower maximum at
higher primary energies is
by measuring the pulse width \cite{ha78}. This technique 
seems to work only at large core distances and at primary energies 
$\ge 10^{15}~eV$
as the two parameters do not show any correlation at  primary energies 
studied here. 

\subsection{Pulse shape parameters}

There are not many results published in the literature regarding the usefulness
of the pulse shape parameters to discriminate hadronic showers from 
electromagnetic showers at lower primary energies. Rodr\'iguez-Fr\'ies 
{\it et al.} \cite{ro}
reported that the pulse width could be used as a discriminating parameter at
primary energies $\ge 100~TeV$ and their results show that sensitivity of
this parameter is too poor at 10 $TeV$. This conclusion is consistent with the
present results.

Patterson \& Hillas \cite{pat89} too have concluded that the pulse widths of 
simulated
proton showers are not very different from those for electromagnetic showers
and hence unlikely to be useful as a discriminating parameter consistent
with the conclusions of the present study. However we find that it could be
a useful parameter for distinguishing inclined hadronic showers.

The present results show that the pulse width and pulse decay time to be 
good discriminating parameters when used with a mask. However our study 
shows no sensitivity to pulse rise time in contrast to the conclusions of
Roberts $et~al.$ \cite{rb98}. Roberts $et~al.$ on the other hand do not use 
fall time to achieve discrimination against hadrons for they find that optimum
quality factor was achieved only by rejecting an excessive fraction of 
$\gamma -$ray cascades, unlike in the present studies.  They also find that
rise time cut discriminates best against heavy primaries while we find that the 
decay time discriminates 
best against heavy primaries. The $\gamma -$ray acceptance fractions when
decay time is used as a discriminating parameter is in the range 30-38\% 
in contrast to the results of Roberts $et~al.$ \cite{rb98}. 

\subsection{Timing jitter}

\begin{figure}
\centerline{\psfig{file=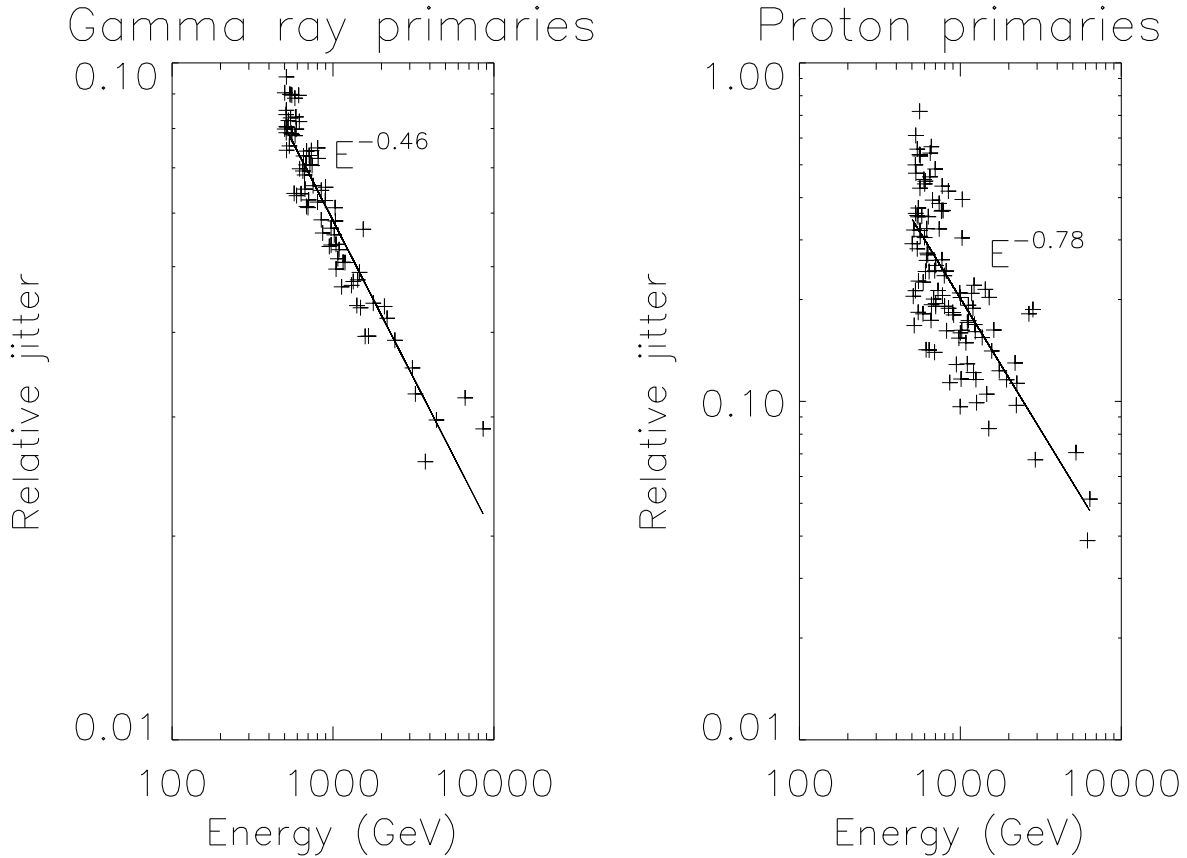,height=12cm}}
\caption{Variation of mean relative jitter for $\gamma -$rays (left)
and protons (right)  as a function of primary energy. 
The primary energies were selected randomly from power law
distribution (differential slope -2.65). The
energy band widths are 500 $GeV$ - 10 $TeV$ for $\gamma -$rays and 500 $GeV$ - 20 $TeV$ 
for
protons. The fitted slopes are indicated for each species.
}
\end{figure}

Cabot {\it et al.} \cite{ca98} have estimated the relative mean times of 
arrival of
\v Cerenkov photons at various core distances. The present results shown in
fig. 3 (shown as +) agree well with theirs. The mean arrival times are fairly
independent of the primary species. However the RMS fluctuations in arrival 
times depend on the range of differential pathlengths which in turn depend 
on the details of the interaction kinematics. Hence they bear the signature
of the primary radiation.

One can ask the question how does the relative jitter vary with primary 
energy.  It could be seen from figures 4a, 4b \& 4d  
that the distributions of relative jitter become narrower with increasing 
energy
for both $\gamma-$ray and proton primaries while average value of the relative
jitter falls with increasing primary energy far more rapidly for proton 
primaries than for $\gamma -$rays.
Figure 5 shows the primary energy dependence of relative jitter averaged over
all core distances.
Hence the quality factor for $\gamma -$ray 
and proton primaries is expected to fall with increasing primary energy as 
could be seen from table 3. 

In an experiment it is not readily obvious how to measure the mean arrival time
and the timing jitter. Normally most of the experiments use discriminators which
trigger at known thresholds. Knowing that a typical \v Cerenkov pulse can be
well described by a lognormal function (modulated by the phototube response)
it is possible to estimate its coefficients knowing when the pulses from 
individual elements cross the preset threshold as well as the pulse height. The
coefficients of the lognormal functions are related to the mean arrival time 
and its RMS value. It is therefore essential to have independent multiple 
sampling of the threshold crossing time and photon density 
from a single \v Cerenkov telescope. Further details are beyond the
scope of this paper.

\section{Conclusions}

Three different measurable parameters based on \v Cerenkov photon arrival 
times at detectors at different core distances at the observational level
have been found to be useful in distinguishing between electromagnetic and 
hadronic showers. The sensitivity of these parameters peaks at around the
hump region and hence best discrimination would be obtained by limiting 
the measurements from detectors around the hump region. 
The sensitivity of these parameters also seems to increase with the
zenith angle of the primary at the top of the atmosphere. This is a useful
property in contrast to the imaging technique which is most sensitive at small
zenith angles \cite{we89}. Similarly, the
use of a circular mask at the focal point also increases the quality factor
in some cases, especially for the curvature of the shower front. However the 
last two conclusions are not true to relative
timing jitter in which case the sensitivity falls with increasing angle of 
incidence with respect to the vertical and marginally by the use of a mask.

\begin{table}
\caption{Table showing the improvement in quality factor when two parameters 
(viz. photon timing jitter and pulse decay time) are used in tandem to 
discriminate against hadronic events. For this 100 showers each of $\gamma -$
rays and protons of energy 500 $GeV$ and 1 $TeV$ respectively as sampled at sea
level are used.}
\vskip 0.3cm
\begin{tabular}{llllll}
\hline
\hline
Parameter & Threshold & Fraction of & Fraction of & Quality \\
Type &  value & $\gamma -$ rays & protons & factor  \\
         & &  accepted (\%)  &  accepted (\%) &  \\
\hline
Pulse Decay    & 3.2 $ns$ & 32.9 & 1.4   & 2.82$\pm$0.07 \\
Time &     &      &      &       \\
&  &&&\\
\hline
Photon Timing &  0.08 & 52.5 &  2.7  & 3.18$\pm$0.06 \\
Jitter &     &      &      &       \\
&   &&&\\
\hline
Decay Time    & 3.2 $ns$ & 26.8 & 0.05  & 12.6$\pm$1.6  \\
\& Timing Jitter & 0.08 &      &      &       \\

\hline
\end{tabular}
\end{table}
 
Although the quality factors for individual parameters are not very large by
themselves, by applying them in tandem would greatly augment the detection
sensitivity of ground based VHE $\gamma -$ray telescopes designed to exploit
the wavefront sampling techniques. Table 17 demonstrates the improvement in
the quality factor when the decay time and the timing jitter are applied
successively to the same sample of showers. The dramatic improvement in the
quality factor exhibits the orthogonal nature of the parameters.
 
The separation efficiency seems to decrease with increasing altitude of 
observation mainly because of the decreasing prominence of the hump in the 
case of electromagnetic showers. This, once again supports the earlier 
conclusion that the \v Cerenkov photons around the hump region are more 
sensitive
to photonic primaries.  Hence sampling of photons from this region of the light
pool is recommended for better discrimination of
hadronic primaries.

\ack{We would like to acknowledge the fruitful discussions with and helpful 
suggestions from Profs. K. Sivaprasad, B. S. Acharya and P. R. Vishwanath 
during the present work.}

\end{document}